\documentclass{appolb}
\pdfoutput=1
\usepackage{graphicx}

\usepackage[preliminary]{nf}

\usepackage[T1]{fontenc}

\usepackage{subcaption}
\usepackage{contour}
\usepackage{siunitx}
\usepackage{ifthen}
\usepackage{mathtools}
\usepackage[capitalize]{cleveref}

\usepackage[backend=bibtex,style=ieee,citestyle=numeric-comp,sorting=none,giveninits=true,url=false]{biblatex}
\AtEveryBibitem{\clearlist{language}}
\usepackage{tikzexternal}
\usepackage{acro}
\acsetup{single}
\DeclareAcronym{FOV}{short=FOV,long=field of view}
\DeclareAcronym{JPET}{short=J\nobreakdash-PET,long=Jagiellonian \acs*{PET}}
\DeclareAcronym{LOR}{short=LOR,long=line of response}
\DeclareAcronym{PET}{short=PET,long=Positron Emission Tomography}
\DeclareAcronym{SiPM}{short=SiPM,long=Sillicon PhotoMultipliers}
\DeclareAcronym{GATE}{short=GATE,long=GEANT4 Application for Tomographic Emission,long-format=\itshape}

\providecommand{\keywords}[1]
{
  \small	
  \textbf{\textit{Keywords---}} #1
}

\begin{document}
\title{Development of the normalization method for the \acl*{JPET} scanner%
}
\author{Aurélien~Coussat~(1,2,*), 
        Wojciech~Krzemien~(3,2), 
        Jakub~Baran~(1,2), 
        Szymon~Parzych~(1,2)
\address{
(1)~Faculty of Physics, Astronomy and Applied Computer Science, Jagiellonian University, 30-348 Krak\'ow, Poland,\\
(2)~Centre for Theranostics, Jagiellonian University, 31-501 Krak\'ow, Poland,\\
(3)~High Energy Physics Division, National Centre for Nuclear Research, Otwock, Swierk, PL-05-400, Poland\\
(*)~\textit{Corresponding author: Aurélien Coussat, aurelien.coussat@uj.edu.pl}
}
\\[3mm]
For the J-PET collaboration
}
\maketitle

\begin{abstract}
This work aims at applying the theory of the component-based normalization for the \acl*{JPET} scanner, currently under development at the Jagiellonian University. In any \acl*{PET} acquisition, efficiency along a line-of-response can vary due to several physical and geometrical effects, leading to severe artifacts in the reconstructed image. To mitigate these effects, a normalization coefficient is applied to each line-of-response, defined as the product of several components. Specificity of the Jagiellonian PET scanner geometry is taken into account. Results obtained from \acs*{GATE} simulations are compared with preliminary results obtained from experimental data.
\end{abstract}


\section{Introduction}
The \ac{JPET} scanner is a high acceptance multi-purpose \ac{PET} detector optimized for the detection of photons from positron-electron annihilation, currently under development at the Jagiellonian University~\cite{raczynski2014,moskal2021positronium,moskal2021testing,niedzwiecki2017}.
The current prototype, named the Modular \ac{JPET}~\cite{moskal2021simulating}, is composed of 24 individual modules arranged cylindrically. Each module is composed of 13 plastic scintillator strips with a size of \qtyproduct[product-units=power]{24x6x500}{\milli\meter}. The scintillators are readout on both sides by a matrix of \ac{SiPM}~\cite{raczynski2014}.

Several effects impact the efficiency of detector strips, such as geometric effects or variation in detector intrinsic efficiencies. To counterbalance the non-uniformity in efficiency, normalization factors can be incorporated into the  image reconstruction procedure.
This contribution is a first step towards proper normalization of the Modular \ac{JPET} scanner. \Cref{sec:mm} describes the normalization factors and how they are computed, \cref{sec:results} shows preliminary results and \cref{sec:conclusion} briefly concludes.

\section{Materials and Methods}
\label{sec:mm}

\subsection{Normalization coefficients}
\label{sec:geom-fact-def}
The proper determination of the normalization coefficient for a given \ac{LOR} permits to compensate for the detector efficiency variation, and for the geometrical effects not included in the projection model. The lack of those corrections leads to artifacts generation and the degradation of the final image quality~\cite{bailey2005}.  
The so called component-based normalization method~\cite{badawi1999} relies on factorization of the normalization coefficients into sub-components that can be estimated separately, and on usage of the fan-sums strategy to lower the variance of the estimations. This work adapts the definitions of \textcite{pepin2011}.

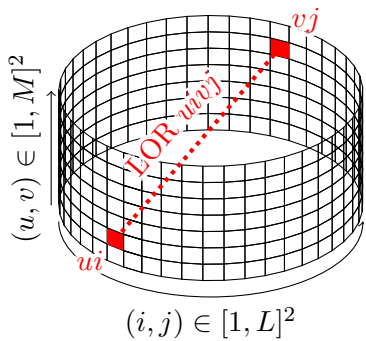
\begin{figure}[htb]
    \centering
    \begin{tikzpicture}[tdplot_main_coords,scale=.5]
        \def\R{4}
        \def\ustep{.5}
        \def\istep{8}
        \def\loru{1}
        \def\lori{352}
        \def\lorv{5}
        \def\lorj{184}
        \def\uspan{6}
        \foreach \u in {0,...,\uspan}
        {
            \foreach \i in {0,\istep,...,360}
            {
                \path[save path=\detector] ({\R*cos(\i)},{\R*sin(\i)},\u*\ustep) coordinate (so)
                    -- ({\R*cos(\i+\istep)},{\R*sin(\i+\istep)},\u*\ustep) 
                    -- +(0,0,\ustep) coordinate (ne)
                    -- ({\R*cos(\i)},{\R*sin(\i)},{(\u+1)*\ustep})
                    -- cycle ;
                \coordinate (center) at ($(so)!.5!(ne)$) ;
                \ifthenelse{\(\u=\loru\AND\i=\lori\)\OR\(\u=\lorv\AND\i=\lorj\)}{
                    \fill[red, use path=\detector] ;
                }{
                    \draw[use path=\detector] ;
                }
                \ifthenelse{\u=\loru\AND\i=\lori}{\coordinate (lorui) at (center) ;}{}
                \ifthenelse{\u=\lorv\AND\i=\lorj}{\coordinate (lorvj) at (center) ;}{}
            }
        }
        \def\Ruv{\R+.2}
        \def\angleuv{-55}
        \draw[->] ({(\Ruv)*cos(\angleuv)},{(\Ruv)*sin(\angleuv)},0)
            -- +(0,0,{(\uspan+1)*\ustep}) node [midway,above,rotate=90] {$(u,v)\in [1,M]^2$} ;
        \tdplotdrawarc[->]{(0,0,-\ustep)}{\R}{\angleuv}{\angleuv+180}{below}{$(i,j)\in [1,L]^2$}
        \draw[dotted,red,ultra thick] (lorui) node [below left] {\contour{white}{$ui$}}
            -- (lorvj) node [above right] {\contour{white}{$vj$}}
            node [midway,above,sloped] {\contour{white}{\acs{LOR} $uivj$}};
    \end{tikzpicture}
    \caption{\Acp{LOR} definition.}
    \label{fig:lors}
\end{figure}

Unlike conventional \ac{PET} scanners, whose detectors are divided into several crystals, the \ac{JPET} scintillator strips are continuous in the axial direction.
We nevertheless define $M$ virtual bins in the axial direction. We also denote as $L$ the number of strips (\num{312} in the case of the Modular \ac{JPET} scanner). The \ac{LOR} that joins portion $u$ of strip $i$ with portion $v$ of strip $j$ is denoted ``\acs{LOR} $uivj$".
These definitions are illustrated in \cref{fig:lors}.

\contourlength{0.15em}

The normalization coefficient for a given \ac{LOR} is given by the product of several normalization factors. Each of these factors accounts for a different effect. The normalization coefficient for the \ac{LOR} $uivj$ is given as~\cite{pepin2011}
\begin{equation}
    \label{eq:eta}
    \eta_{uivj}=b\ax_u\cdot b\ax_v\cdot g\ax_{uv}\cdot g\tr_{ij}\cdot f\tr_{ij}\cdot\epsilon_{ui}\cdot\epsilon_{vj}
\end{equation}
where $b\ax$ represents the axial block profile factors, $g\ax$ the axial geometric factors, $g\tr$ the transverse geometric factors, $f\tr$ the transverse interference function and $\epsilon$ the intrinsic detector efficiencies.
Note that transverse interference functions are designed to compensate for non-uniformity of detection efficiency with respect to the location of a crystal in a detector block, and can be ignored in the context of the \ac{JPET} scanner due to the design of its detectors.

Axial block profile factors, axial geometric factors, and intrinsic detector efficiencies are computed from the acquisition of a uniform cylindrical source centered on the scanner axis, and we denote as $\tcyl_{uivj}$ the number of true coincidences measured along \ac{LOR} $uivj$ during the acquisition.
Transverse geometric factors are computed from the acquisition of a uniform annular source, and we denote as $\tann_{uivj}$ the number of true coincidences measured along \ac{LOR} $uivj$ during the acquisition.
``True coincidences'' refer here to coincidences that have not undergone any scattering (in the phantom or in the detector) and that are not accidental.

Axial block profile factors $b\ax$ normalize true coincidences along axial planes, that is the planes comprising the \acs{LOR} whose detectors are located at the same axial position ($u=v$). They are defined as
\begin{equation}
    b\ax_{u}=\sqrt{\frac{\frac{1}{M}\sum_{v=1}^M\sum_{i=1}^L\sum_{j=1}^L\tcyl_{vivj}}
        {\sum_{i=1}^L\sum_{j=1}^L\tcyl_{uiuj}}}.
\end{equation}

Axial geometric factors account for efficiency variations caused by the detector geometry in the axial direction.
They are defined between two axial positions $u$ and $v$ as
\begin{equation}
    \label{eq:g_ax}
    g\ax_{uv}=\frac{
        \frac{1}{M^2}
        \sum_{u'=1}^M\sum_{v'=1}^Mb\ax_{u'}\cdot
        b\ax_{v'}\sum_{i=1}^L\sum_{j=1}^L\tcyl_{u'iv'j}
        \cos\theta
    }{
        b\ax_{u}\cdot
        b\ax_{v}\sum_{i=1}^L\sum_{j=1}^L\tcyl_{uivj}
        \cos\theta
    }
\end{equation}
where $\theta$ is the angle between the \ac{LOR} and the transverse plane.

Transverse geometric factors also account for efficiency variations caused by the detector geometry, but this time along transverse planes. They are defined for a radial distance $r\in[1;K]$, where $K$ is the number of radial bins, as
\begin{equation}
    g\tr_r=\frac{
        \frac{1}{K}\sum_{r'=1}^K\sum_{u=1}^M\sum_{v=1}^M\underset{x_r(i,j)=r'}{\sum_{i=1}^L\sum_{j=1}^L}\cann_{vivj}
    }{
        \sum_{u=1}^M\sum_{v=1}^M\underset{x_r(i,j)=r}{\sum_{i=1}^L\sum_{j=1}^L}\cann_{vivj}
    }
\end{equation}
where $x_r(i,j)$ represents the radial distance for the \ac{LOR} joining strips $i$ and $j$, and where $\cann_{uivj}$ represents the number of coincidence for \ac{LOR} $uivj$ with the correction given by
$
    \cann_{uivj}=a_{uivj}\cdot b\ax_u\cdot b\ax_v\cdot g\ax_{uv}\cdot\epsilon_{ui}\cdot\epsilon_{vj}\cdot\tann_{uivj}.
$
Here, $a_{uivj}$ corresponds to the inverse of the analytical projection of the source.

Intrinsic detector efficiency $\epsilon_{ui}$ represents the ability of the strip portion located at ring $u$ and strip $i$ to convert gamma photons into light. This parameter is computed using the fan-sum algorithm as
\begin{equation}
    \epsilon_{ui}=\frac{
        \frac{1}{L}\sum_{i'=1}^L\sum_{v=1}^M\sum_{j=1}^L\tcyl_{ui'vj}
    }{
        \sum_{v=1}^M\sum_{j=1}^L\tcyl_{uivj}
    }.
\end{equation}

\subsection{Data acquisition}
\label{sec:data-acquisition}

\begin{figure}[htb]
    \centering
    \begin{subfigure}{.3\textwidth}
        \includegraphics[width=\textwidth]{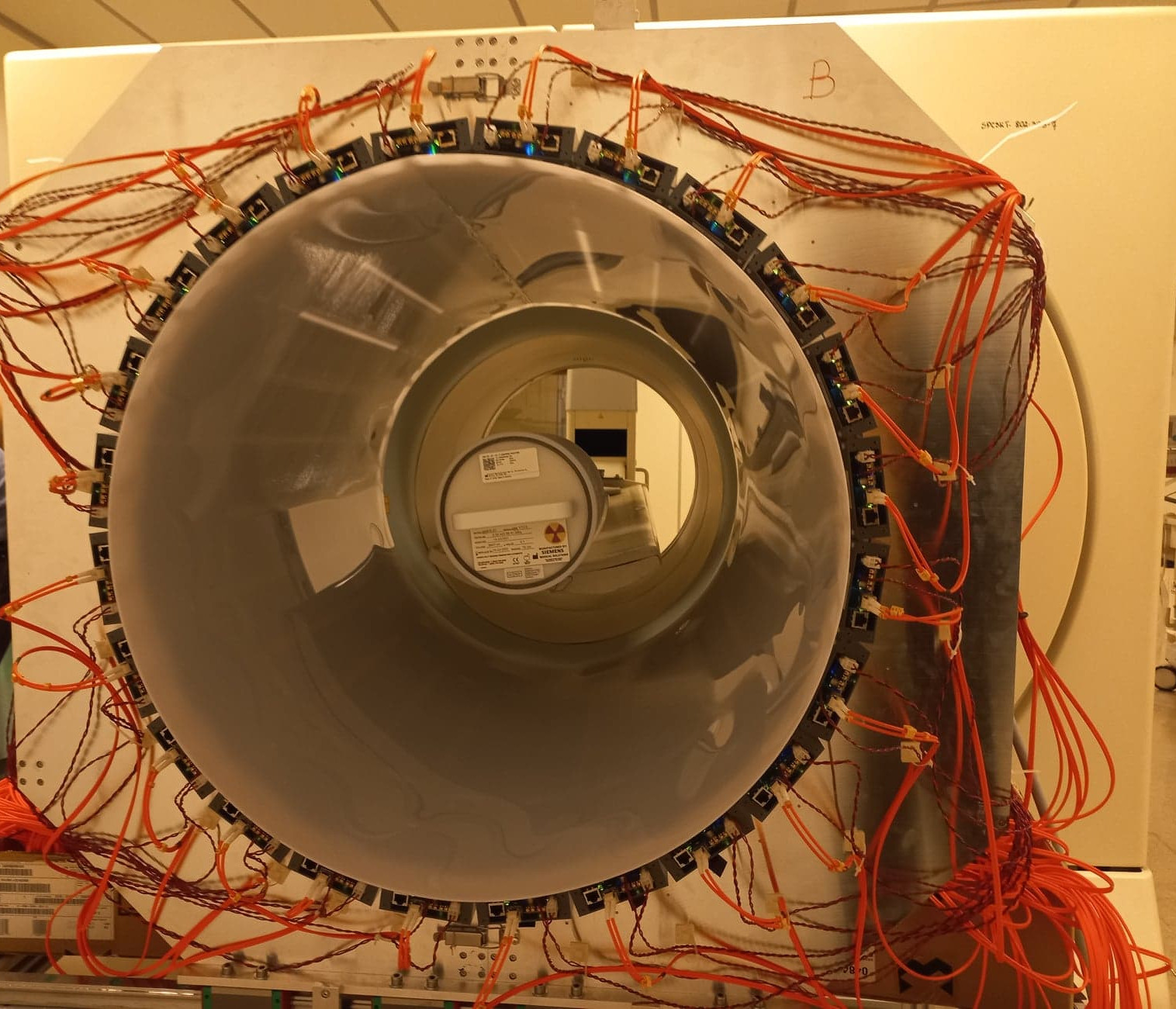}
        \caption{Data acquisition from a cylindrical phantom.}
        \label{fig:data_warsaw}
    \end{subfigure}
    \hfill
    \begin{subfigure}{.3\textwidth}
        \includegraphics[width=\textwidth]{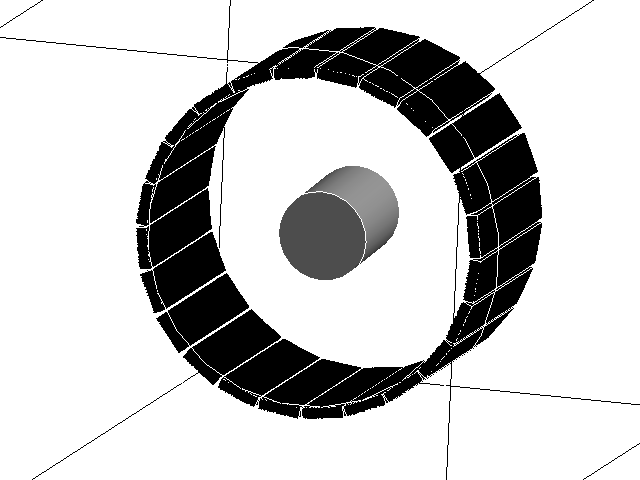}
        \caption{Monte Carlo simulation of a uniform cylindrical phantom.}
        \label{fig:data_cylinder}
    \end{subfigure}
    \hfill
    \begin{subfigure}{.3\textwidth}
        \includegraphics[width=\textwidth]{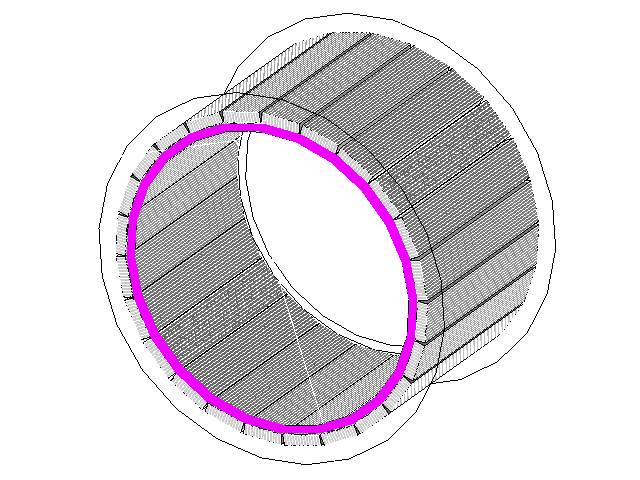}
        \caption{Monte Carlo simulation of a uniform annular phantom (in pink, not to scale).}
        \label{fig:data_annulus}
    \end{subfigure}
    \caption{Data acquisition setups.}
    \label{fig:data}
\end{figure}

In order to compute the normalization factors, the acquisition of a Siemens CS-27 cylindrical phantom was performed. The cylinder was \SI{50}{\centi\meter} long, had a radius of \SI{10}{\centi\meter}, and a capacity of \SI{8407}{\milli\liter}. The cylinder was filled with \SI{88.43}{\mega\becquerel} of Gallium-68 and placed at the center of the Modular \ac{JPET}. The setup is shown in \cref{fig:data_warsaw}.
Note that we have not performed any coincidence filtering in this case. At this stage of development, we consider the effect of scattered and accidental coincidences as negligible, and we leave their filtering for future works.

Two simulations of both a cylindrical and an annular phantom were also performed using the \acf{GATE}~\cite{sarrut2021}.
The cylindrical setup simulates a \SI{1800}{\second} acquisition of the cylinder described above, with an activity of \SI{100}{\mega\becquerel}, placed at the center of the detector. The cylindrical simulation setup is illustrated by \cref{fig:data_cylinder}.
The annular simulation was performed using a moving ring source. The ring source was \SI{1}{\centi\meter} thick and \SI{2.5}{\milli\meter} long with \SI{10}{\mega\becquerel} of activity. A number of 200 positions were axially simulated, each simulating \SI{100}{\second} of acquisition, resulting in a total time of \SI{20000}{\second}. The annular simulation setup is illustrated by \cref{fig:data_annulus}.
In both simulations, the scattered and accidental coincidences were completely filtered out based on recorded hit data, resulting in \num[scientific-notation=true,round-precision=2,round-mode=places]{50759135} true coincidences (out of \num[scientific-notation=true,round-precision=2,round-mode=places]{87673718} coincidences, \SI[round-precision=2,round-mode=places]{57.895497257228214}{\percent}) for the cylindrical phantom, and \num[scientific-notation=true,round-precision=2,round-mode=places]{89555439} true coincidences (out of \num[scientific-notation=true,round-precision=2,round-mode=places]{119543736} coincidences, \SI[round-precision=2,round-mode=places]{74.91437192493298}{\percent}) for the annular phantom.

Values used for the various parameters described in \cref{sec:geom-fact-def} are the following: $M=\num{25}$, $L=\num{312}$, $K=\num{25}$.


\section{Preliminary results}
\label{sec:results}

\begin{figure}[htb]
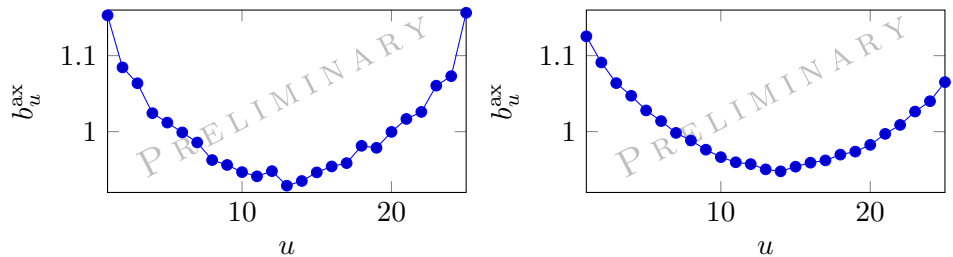

    \centering
    \begin{subfigure}{.5\textwidth}
        \plotbax{data/warsaw_mc/b_ax.dat}
        \caption{Computed from Monte Carlo data.}
        \label{fig:b_ax_mc}
    \end{subfigure}%
    \begin{subfigure}{.5\textwidth}
        \plotbax{data/warsaw/b_ax.dat}
        \caption{Computed from cylinder acquisition.}
        \label{fig:b_ax_warsaw}
    \end{subfigure}
    \caption{Axial block profile factors ($b\ax$). Lines between data points illustrate the trend but do not refer to data interpolation.}
    \label{fig:b_ax}
\end{figure}

\Cref{fig:b_ax} shows the axial geometric factors $b\ax$. The lower values near the axial center of the scanner denote a higher detection efficiency. The results from Monte Carlo data (\cref{fig:b_ax_mc}) display stronger fluctuations than those from real data (\cref{fig:b_ax_warsaw}) probably due to the lower statistics of the Monte Carlo sample. On the other hand, the factors obtained with real data are asymmetric: we suppose that this is due to the cylinder being slightly tilted by about \ang{3} during the acquisition. Further investigations are required to conclude on this observation.

\begin{figure}[htb]
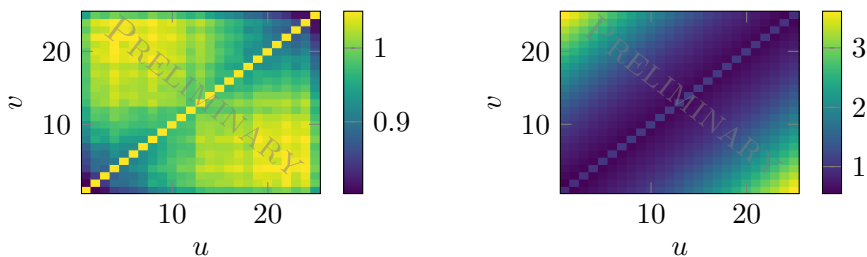

    \centering
    \begin{subfigure}{.5\textwidth}
        \plotgax[][1.05]{data/warsaw_mc/g_ax.dat}
        \caption{Computed from Monte Carlo data.}
        \label{fig:g_ax_mc}
    \end{subfigure}%
    \begin{subfigure}{.5\textwidth}
        \plotgax{data/warsaw/g_ax.dat}
        \caption{Computed from cylinder acquisition.}
        \label{fig:g_ax_warsaw}
    \end{subfigure}
    \caption{Axial geometric factors ($g\ax$).}
    \label{fig:g_ax}
\end{figure}

\Cref{fig:g_ax} shows the axial geometric factors $g\ax$. As expected from the definition of the axial geometric factors (\cref{eq:g_ax}), the factors are constant along the diagonal, that is where $u=v$. At the edges, where the ring difference is large, the efficiency is lower due to the decreased probability of the \acp{LOR}, hence the higher values of the axial geometric factors. Note the difference in color scale between \cref{fig:g_ax_mc} and \cref{fig:g_ax_warsaw}: we suggest that these are due to \ac{LOR} obliqueness and the coincidence filtering strategy applied by the \ac{JPET} scanner with respect to ring difference. However, further investigations are needed to validate our hypothesis.

\begin{figure}[htb]
    \centering
    \begin{subfigure}{.5\textwidth}
        \plotepsilon{data/warsaw_mc/epsilon.dat}
        \caption{Computed from Monte Carlo data.}
        \label{fig:espilon_mc}
    \end{subfigure}%
    \begin{subfigure}{.5\textwidth}
        \plotepsilon{data/warsaw/epsilon.dat}
        \caption{Computed from cylinder acquisition.}
        \label{fig:epsilon_warsaw}
    \end{subfigure}
    \begin{subfigure}{\textwidth}
        \begin{tikzpicture}
            \begin{axis}[
                xlabel=$i$,
                ylabel=Average of $\epsilon_{ui}$,
                xmin=1,
                xmax=312
              ]
              \addplot[only marks] table[x index=0,y index=1] {data/warsaw/epsilon_integrated.dat};
              \preliminary*
            \end{axis}
        \end{tikzpicture}
        \caption{Average value of $\epsilon_{ui}$ for each strip $i$.}
        \label{fig:epsilon_warsaw_integrated}
    \end{subfigure}
    \caption{Intrinsic detector efficiencies ($\epsilon$).}
    \label{fig:epsilon}
\end{figure}
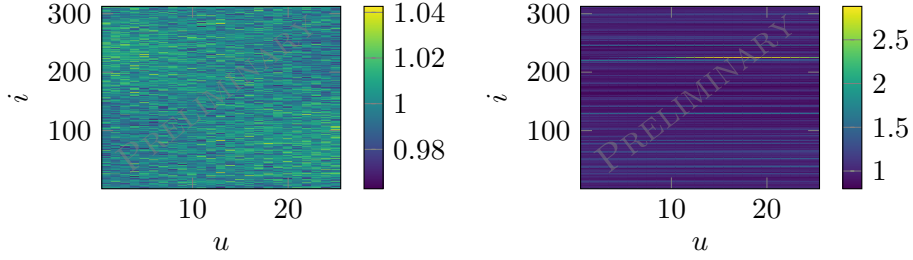
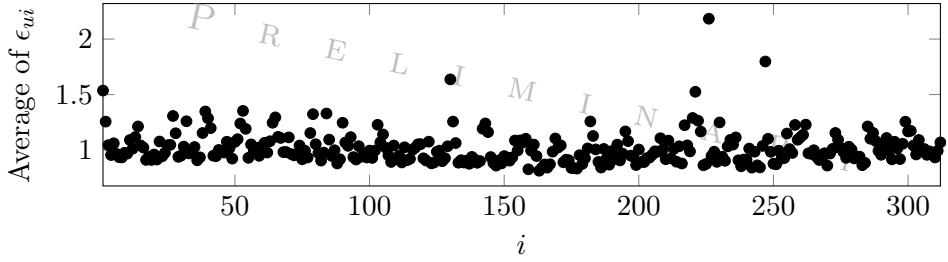

\Cref{fig:epsilon} shows the intrinsic detector efficiencies $\epsilon$. In \cref{fig:espilon_mc}, as we considered a \ac{GATE} simulation with perfect detectors and uniform efficiencies, the factors are uniform and the small variations that appear are entirely due to statistical noise. On the other hand, \cref{fig:epsilon_warsaw} highlights which areas of the detectors have a lower efficiency. \Cref{fig:epsilon_warsaw_integrated} averages on a strip-basis the values presented in \cref{fig:epsilon_warsaw}, which makes the anomalies clearly appear.

\begin{figure}[htb]
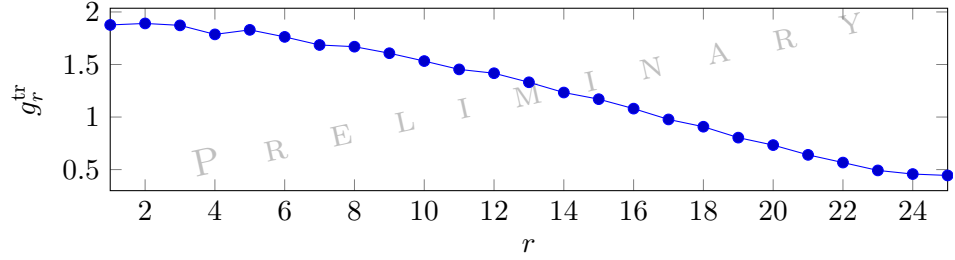

\centerline{%
\plotgtr{data/mc/g_tr.dat}}
\caption{Transverse geometric factors ($g\tr$). Lines between data points illustrate the trend but do not refer to data interpolation. This result is Monte Carlo-based only due to the lack of a dedicated measurement.}
\label{fig:g_tr}
\end{figure}

\Cref{fig:g_tr} shows the transverse geometric factors $g\tr$. The low values for large radial distances show that efficiency is higher near the edge of the \ac{FOV}, as expected from the geometry of the detector strips, due to the \ac{LOR} obliqueness.

\section{Discussion and conclusions}
\label{sec:conclusion}

Normalization components highlight the relative importance of several physical and geometrical effects. They can be used to obtain insights of the efficiency of different aspects of the scanner, such as the efficiency of the detectors, or the scanner response with respect to \ac{LOR} obliqueness.
Due to the design of the \ac{JPET} scanner, where the detector strips are axially continuous, the definition of some normalization factors must be adapted.
Future work consists of interpolating the normalization factors that have an axial dependency in order to compute normalization factors for any point along the whole strip, and to assess the improvements in image quality when taking into account all the normalization factors during image reconstruction.
The final goal is to apply the same procedure to the future Total-Body \ac{JPET} scanner~\cite{pet_clinics,alavi2021}.

\section{Acknowledments}
The authors acknowledge  the support provided by 
the Foundation for Polish Science through the
TEAM POIR.04.04.00-00-4204/17 program; 
the National Science Centre of Poland through grants MAESTRO no.
2021/42/\linebreak[0]A/ST2/00423 and OPUS no. 2019/35/B/ST2/03562;
the Ministry of Education and Science through grant no. SPUB/SP/490528/2021;
the SciMat and qLIFE Priority Research Areas  budget under the program {\it Excellence Initative - Research University} at the Jagiellonian University,
and Jagiellonian University project no. CRP/0641.221.2020.

\printbibliography

\end{document}